# Breakdown of compensation and persistence of non-saturating magnetoresistance in WTe$_2$ thin flakes

Yilin Wang,[1,3] Kefeng Wang,[1] Janice Reutt-Robey,[3] Johnpierre Paglione[1] and Michael S. Fuhrer[1,2*]

[1]*Center for Nanophysics and Advanced Materials, University of Maryland, College Park, MD 20742, USA*

[2]*School of Physics, Monash University, Victoria 3800, Australia*

[3]*Department of Chemistry and Biochemistry, University of Maryland, College Park, MD 20742, USA*

* michael.fuhrer@monash.edu

We present a detailed study of magnetoresistance $\rho_{xx}(H)$, Hall effect $\rho_{xy}(H)$, and electrolyte gating effect in thin (<100 nm) exfoliated crystals of WTe$_2$. We observe quantum oscillations in $H$ of both $\rho_{xx}(H)$ and $\rho_{xy}(H)$, and identify four oscillation frequencies consistent with previous reports in thick crystals. $\rho_{xy}(H)$ is linear in $H$ at low $H$ consistent with near-perfect electron-hole compensation, however becomes nonlinear and changes sign with increasing $H$, implying a breakdown of compensation. A field-dependent ratio of carrier concentrations $p/n$ can consistently explain $\rho_{xx}(H)$ and $\rho_{xy}(H)$ within a two-fluid model. We also employ an electrolytic gate to highly electron-dope WTe$_2$ with Li. The non-saturating $\rho_{xx}(H)$ persists to $H$ = 14 T with magnetoresistance ratio exceeding 2 x 10$^4$ %, even with significant deviation from perfect electron-hole compensation ($p/n$ = 0.84), where the two-fluid model predicts a saturating $\rho_{xx}(H)$. Our results suggest electron-hole compensation is not the mechanism for extremely large magnetoresistance in WTe$_2$, other alternative explanations need to be considered.

After the successful exfoliation of single layer graphene in 2004, two-dimensional layered materials, such as transition-metal dichalcogenides, are under intense study due to their excellent electrical, optical, thermal and mechanical properties[1]. Among TMDCs, Semi-metal tungsten ditelluride ($WTe_2$) has attracted significant research interest since the recent discovery of its extremely large, non-saturating magnetoresistance (XMR)[2], which could find potential applications in magnetic memory and sensing. The XMR in $WTe_2$ is believed to arise from perfect compensation of electron ($n$) and hole ($p$) carrier densities[2], similar to the case observed in high purity graphite and bismuth[3]; this is supported by angle-resolved photoemission spectroscopy (ARPES) experiments, which observed nearly identical sized electron and hole pockets[4]. These results have been complicated by further ARPES study[5] which revealed a more complicated Fermi surface, and transport studies of Shubnikov-de-Haas (SdH) quantum oscillations[6, 7], which reflect the existence of four Fermi pockets. The application of pressure, which increases the difference between electron- and hole- pockets, suppresses the magnetoresistance[8, 9, 10], further supporting the electron-hole compensation scenario. However the origin of XMR in $WTe_2$ is still unclear and remains under discussion[5, 7], therefore more experiments to measure[11] and tune the degree of compensation are highly desired to clarify this important issue.

In this work, we present detailed temperature and gate-dependent magnetoresistivity $\rho_{xx}(H)$ and Hall resistivity $\rho_{xy}(H)$ measurements on exfoliated $WTe_2$ thin (<100 nm) flakes, in an attempt to reveal the relationship between magnetoresistivity and electron and hole concentrations. We observe quantum oscillations in both $\rho_{xx}(H)$ and $\rho_{xy}(H)$, with four oscillation frequencies resolved in the fast Fourier transform (FFT) analysis, consistent with earlier reports[6, 7]. Hall resistivity exhibits pronounced temperature dependent nonlinear behavior and changes its sign under magnetic field; the non-linearity is strong evidence of a breakdown in the perfect electron-hole compensation. By fitting the Hall resistivity with two-fluid model, we found the two carriers are well compensated under low field where $\rho_{xy}(H)$ is linear in $H$; while they are not compensated under high field: holes have the larger carrier population and electrons show higher mobility. We also tuned the doping and compensation ratio *p/n* through electrolyte gating. We find that the large non-saturating magnetoresistivity persists, up to 2 x 10$^4$ % at $H$ = 14 T, even though the electrons and holes become significantly uncompensated, with *p/n* as small as 0.84. For such uncompensated samples, a two-fluid model predicts a saturating magnetoresistivity not seen in experiment. The results indicate that near-perfect electron-hole compensation is not required for non-saturating magnetoresistance and suggests alternative explanations need to be explored.

**Results**

**Magnetoresistance in WTe$_2$ thin flake.** Figure 1(a), inset shows a typical optical image of one of our WTe$_2$ devices with a thickness of about 70 nm. WTe$_2$ is found to have a distorted 1T structure with tungsten chains sandwiched between two layers of tellurium atoms, which makes it cleave easily into atomically thin layers[12]. Figure 1 shows the temperature-dependent resistivity $\rho_{xx}(T)$ of this device at $H$ = 0; the resistivity decreases monotonically with decreasing temperature, exhibiting metallic behavior. When an external magnetic field $H \parallel c$ is applied, the resistivity increases monotonically with increasing magnetic field, no sign of saturation up to 14 T was observed, as shown in Fig. 1(b), which presents field-dependent resistivity $\rho_{xx}(H)$ at various temperatures, clear SdH quantum oscillations are observed below 10 K. These results are similar to those reported for bulk single crystals[2, 6, 7, 8, 13]. Figure 1(c) shows the magnetoresistance ratio (*MR*), $MR = [\rho(H) - \rho(0)]/\rho(0)$, as a function of magnetic field in a log-log plot at various temperatures, the linear dependence indicates power-law $MR \sim H^n$ behavior with $n = 1.6$ at $T = 2$ K, and gradually increasing to $n = 1.9$ at $T = 100$ K[14].

**Nonlinear Hall effect.** Figure 2 shows the field-dependent Hall resistivity $\rho_{xy}(H)$ measured at various temperatures. At temperatures below 50 K, $\rho_{xy}(H)$ exhibits nonlinear behavior; at temperatures above 100 K, $\rho_{xy}(H)$ behavior is quite linear. At low field, $\rho_{xy}$ is positive with a value that depends nearly linearly on field; with increasing field, the slope gradually changes from positive to negative and $\rho_{xy}$ changes its sign. With increasing temperature, the field where $\rho_{xy}$ changes the sign continuously increases. (The nonlinear behavior of Hall resistivity was confirmed in several devices, see Supplementary Fig.1) These features are similar to that of other XMR materials, such as NbSb$_2$[15], and definitely indicate multiple Fermi pockets; at least two types of carriers play a role in the transport properties of WTe$_2$. Notably, the non-linearity of $\rho_{xy}$ within a two-fluid model occurs only if the two carrier densities are unequal.[15, 16] Detailed analysis of $\rho_{xy}(H)$ will be presented below. At 2 K and high field (> 5 T), clear quantum oscillations superimposed on the smooth background were observed.

**SdH quantum oscillations and quantum Hall oscillations.** Quantum oscillations of magnetoresistivity and Hall resistivity are obtained by subtracting the smoothly varying background fitted to a polynomial and are shown in Fig. 3(a) at various temperatures. The amplitude of the quantum oscillations of both $\rho_{xx}(H)$ and $\rho_{xy}(H)$ decreases with increasing $T$, and the oscillations are not observed for $T >15$ K. In order to resolve the frequency of quantum oscillations, a Fourier transform analysis was carried out on $\rho_{xx}$ vs. $1/H$ with results shown in Fig. 3(b). Four major peaks at frequencies $F_{\alpha 1} = 78$ T, $F_{\beta 1} = 105$ T, $F_{\beta 2} = 124$ T and $F_{\alpha 1} = 151$ T were identified, which may correspond to four distinct Fermi pockets ($F_\alpha$ and $F_\beta$ correspond to previously observed hole- and electron-pockets, respectively)[6, 7]. Fourier transform analysis of $\rho_{xy}$ vs. $1/H$ (red curve in Fig. 3b) shows similar oscillation frequencies as the SdH oscillations. The observation of SdH oscillations with accompanying quantum Hall oscillations suggests the Landau levels in this system are well resolved, indicating the high quality of the device.

**Two-fluid model.** In order to further analyze $\rho_{xx}(H)$ and $\rho_{xy}(H)$ in WTe$_2$, we adopt a two-fluid model[17]:

$$\rho_{xx} = \frac{1}{e}\frac{(n\mu + p\mu') + (p\mu + n\mu')\mu\mu' H^2}{(n\mu + p\mu')^2 + [(p-n)\mu\mu' H]^2} \tag{1}$$

$$\rho_{xy} = \frac{1}{e}\frac{(n\mu^2 - p\mu'^2) - (p-n)\mu^2\mu'^2 H^2}{(n\mu + p\mu')^2 + [(p-n)\mu\mu' H]^2} H \tag{2}$$

, where $n$ ($p$) and $\mu$ ($\mu'$) are the carrier density and mobility for electrons (holes),

respectively, $H$ is the magnetic field. The $\rho_{xy}(H)$ at low-field (< 4 T) is linear (slope independent of $H$), indicating $p = n$, the electrons and holes are well compensated at low field. The slope of $\rho_{xy}(H)$ at high-field is nonlinear and strongly field dependent, which is inconsistent with Eqns. (1) and (2) with $p = n$; the two-fluid model cannot explain $\rho_{xy}(H)$ with fixed parameters $n$, $p$, $\mu$, $\mu'$. Indeed, recent DFT calculations predict the Fermi surface of WTe$_2$ exhibit a strong magnetic field dependent behavior[7]. Since the non-linearity of $\rho_{xy}(H)$ is highly suggestive that electrons and holes are not well compensated, i.e. $p \neq n$, the simplest modification to the two-fluid model is to adopt a dependence of the ratio $p/n$ on $H$ as follows. The combined linear fitting of $\rho_{xy}(H)$ and second-order polynomial fitting of $\rho_{xx}(H)$ at low field (< 4 T) with the two-fluid model Eqns. (1) and (2) with $p = n$ yields mobility ($\mu$ and $\mu'$) and carrier densities $p = n$. Figure 4(a) shows the extracted mobilities and carrier densities at various temperatures; at 2 K, $p = n = 2.45 \times 10^{19}$ cm$^{-3}$, $\mu = 0.77$ m$^2$/Vs and $\mu' = 0.40$ m$^2$/Vs, and $p$, $n$, $\mu$, $\mu'$ gradually decrease with increasing temperature. We then assume that the non-linearity of $\rho_{xy}(H)$ for H > 4 T is attributed entirely to the variation of the $p/n$ ratio; we fix $\mu$, $\mu'$, and $p + n$ by their low-field values. Figure 4(b) shows the $p/n$ ratio (obtained by inverting Eqn. (2) and holding $p + n$, $\mu$ and $\mu'$ constant) as a function of magnetic field at various temperatures. The $p/n$ ratio varies continuously with field and gives consistent results at every temperature from 2 K $\leq T \leq$ 35 K. As a check we use the $p/n$ ratio in Fig. 4b together with $\mu$, $\mu'$, and $p + n$ to predict $\rho_{xx}(H)$ using Eqn. (1) as shown in Fig. 4c. The predicted $\rho_{xx}(H)$ is in reasonable agreement with experiment. We note that better agreement could be obtained by allowing $\mu$, $\mu'$, and/or $p + n$ to vary as well. The variation of $\rho_{xy}$ with $H$ could in principle be explained with $p = n$ however it would require e.g. drastic variation including change in relative magnitude in $\mu$ and $\mu'$, which we find much less likely especially in a scenario that conserved $p = n$. Hence the non-linear $\rho_{xy}(H)$ provides strong evidence that near-perfect compensation does not survive to high $H$. The $p/n$ ratio is slightly above 1 below 50 K at $H > 4$ T, indicating the concentration of holes is higher than that of electrons; the mobility of holes is smaller (~ 50%) than that of electrons. The higher carrier population and lower mobility of holes give rise to the sign-reversal of $\rho_{xy}$ in magnetic field at low temperature. The carrier mobility is above the threshold value for observation of quantum oscillations in magnetoresistance and Hall resistivity: $\mu B \approx 1$. At 100 K, we find a $p/n$ ratio smaller than 1, indicating the concentration of electrons is higher than that of holes, which is consistent with the observation via ARPES experiments of increased electron doping due to thermal activation above 100 K[4] (albeit at $H = 0$).

**Electrolytic gating of WTe$_2$.** We also used an electrolyte gate to tune the absolute and relative densities of electrons and holes in WTe$_2$ thin crystals. Figure 5 shows the Hall resistivity and magnetoresistivity of a second WTe$_2$ device with the thickness of ~ 70 nm, tuned by electrolyte gating (LiClO$_4$ and PEO; see Methods)[18]. As shown in Fig. 5 (a), the $\rho_{xy}(H)$ curve at V$_g$ = 0 V is highly non-linear, with holes dominating at high field. After applying a positive gate voltage, the $\rho_{xy}(H)$ is uniformly positive (electron-like). Applying a negative gate voltage had little effect. The electron-doping at positive gate voltage is irreversible, persisting after returning the device to V$_g$ = 0 V

(see Supplementary Fig.2). The observations are consistent with irreversible doping due to intercalation of donor Li ions into the spacing layer between Te atoms[19] at positive gate voltage. The Hall resistivity at $V_g = 3$ V can be well fit with Eqn. (2) with the constraint of $\rho_{xx}(0) = \dfrac{1}{e(n\mu + p\mu')}$ yielding $n_e = 1.36 \times 10^{20}$ cm$^{-3}$ and $p/n$=0.84, suggesting WTe$_2$ has become strongly electron-doped. Figure 5(b) presents $\rho_{xx}(H)$ at various gate voltage, the $\rho_{xx}$ value decreases with increasing doping. Interestingly, large magnetoresistance was also observed in electron-doped WTe$_2$ with no evidence of saturation up to 14 T; $\rho_{xx}(H)$ is never sub-linear in $H$. While $\rho_{xx}(H = 14$ T) is lower at $V_g = 3$ V (high electron doping), compared to $V_g = 0$ V, it should be noted that $\rho_{xx}(0)$ also decreases when increasing doping, as shown in the inset of Fig. 5(b). Figure 5(c) shows the field dependence of MR at various gate voltages, the MR is slightly gate dependent and is in fact even larger in the electron-doped WTe$_2$ compared to $V_g = 0$ V over the measured field range of <14 T (see Supplementary Fig.2). In the case of graphite and bismuth, MR tends to saturate when B exceeds several Tesla because of the slight deviation of perfect compensation of electrons and holes[20, 21]. Notably the $p/n$ ratio different than 1 at $V_g = 3$ V predicts a saturating MR, as shown in the blue dashed line in Figs. 5(b) and (c), which is not observed in the experiment. The large MR in WTe$_2$ persists even when electrons and holes are not compensated, as determined by the Hall effect and electrolyte gating effect, which suggest that perfect electron-hole compensation is not be necessary for the XMR in WTe$_2$, and other mechanisms should be considered[5].

**Metal-insulator crossover at high magnetic field.** When a field is applied, $\rho_{xx}(T)$

exhibits insulating behavior below the 'turn on' temperature $T^*$: at T < $T^*$, $\rho_{xx}$ increases markedly with decreasing temperature; at T > $T^*$, $\rho_{xx}$ decreases with decreasing temperature, as shown in the red curve in Fig. 6. Similar $T^*$ behavior is also observed in other XMR materials[15, 22, 23, 24, 25] and its origin is still under debate. The black curve in Fig. 6 shows the temperature dependence of Hall coefficient, $R_H$, at a field of 14 T. When T > $T^*$, $R_H$ changes slowly and remains negative, illustrating the transport properties are dominated by electrons at high temperature; when T < $T^*$, $R_H$ increases significantly, crosses zero and becomes positive, suggesting the carrier of holes dominates the transport properties at low temperature. The remarkable changes in $R_H$ and $\rho_{xx}$ occur simultaneously, indicating both of them may arise from the same mechanism. The $T^*$ behavior is commonly attributed to a magnetic-field-driven metal-insulator transition[22, 26], while the observation of continuous increase of $R_H$ with decreasing temperature is inconsistent with the existence of insulating gap, which should result in decrease of $R_H$ with decreasing temperature. We also observed similar $T^*$ behavior in strongly electron-doped $WTe_2$ under electrolyte gating (see Supplementary Fig.3). The strong $T$ dependence and the sign-reversal of $R_H$ suggests the Fermi surface of $WTe_2$ differs with temperature and the observed $T^*$ behavior may originate from a change of electronic structure with temperature.

**Methods**

**Device fabrication and experimental details of measurements.** $WTe_2$ thin flakes are obtained by mechanical exfoliation of ribbon-like single crystals on a 300-nm-$SiO_2$/Si substrate. Single crystal platelets of WTe$_2$ were grown via the high-temperature self-flux method. High purity elemental W and Te in W/Te = 1/49 ration were placed in the alumina crucible which was sealed in quartz tube. The sealed tube was heated to 1000 °C and then slowly cooled down to 480 °C where the residual Te flux was decanted with a centrifuge. The electrical contacts are defined with standard electron beam lithography and thermally evaporated Cr/Au (5 nm/100 nm). The Hall bar geometry is defined via SF$_6$ plasma etching. Four-probe measurements of magnetoresistivity and Hall resistivity were conducted by lock-in techniques at a low frequency of 3.7 Hz and in a commercial Quantum Design PPMS-14. The Hall voltage was recorded in both polarities of the magnetic field and anti-symmetrized to remove longitudinal voltage components. Polymer electrolyte was prepared by dissolving the mixture of LiClO$_4$ and polyethylene oxide (PEO) in the weight ratio 0.12:1 in methanol and then stirring overnight at room temperature[18].

## Acknowledgments

Materials synthesis efforts were supported by the Gordon and Betty Moore Foundation Grant No. GBMF4419. M.S.F. is supported by the Australian Research Council.

## Author contributions

Y.W., J.R.R. and M.S.F. conceived the experiments. Y.W. fabricated devices and performed the measurements. K.W and J.P prepared the bulk crystals of WTe$_2$. Y.W. and M.S.F. analyzed the data. Y.W. prepared the manuscript draft and all authors contributed to the editing of the manuscript.

## Additional Information

**Competing financial interests**: The authors declare no competing financial interests.

Figure Captions

**Figure 1 | Temperature and field dependence of magnetoresistance of WTe$_2$.** (a) Temperature dependent resistivity $\rho_{xx}$ of WTe$_2$ thin flakes at zero magnetic field. The inset shows the typical optical image of our WTe$_2$ devices with Hall-bar geometry. (b) Magnetic field dependence of resistivity $\rho_{xx}(H)$ at various temperatures. (c) Magnetoresistance ratio (MR) versus field in a log-log plot; straight lines indicate a power-law relationship $MR \sim H^n$.

**Figure 2 | Temperature and field dependence of Hall resistivity of WTe$_2$.** The dependence of the Hall resistivity on magnetic field is shown at various temperatures indicated in the legend. The inset shows the same data at low magnetic fields.

**Figure 3 | Shubnikov-de Haas (SdH) quantum oscillations and quantum Hall oscillations.** (a) Quantum oscillations of the longitudinal resistivity (SdH, upper panel) and Hall resistivity (lower panel) after subtraction of a polynomial background at various temperatures. (b) Fast Fourier transform of the quantum oscillations of $\rho_{xx}$ and $\rho_{xy}$ vs. $1/H$ at $T = 2$ K; four oscillation frequencies are resolved as denoted on the plot.

**Figure 4 | Fitting the nonlinear Hall resistivity of WTe$_2$ with two-fluid model.** (a) The mobilities μ, μ' and carrier densities $p = n$ of electrons and holes at various temperatures, extracted from data in Fig. 1(b) and Fig. 2 at $H < 4$ T. (b) The $p/n$ ratio as a function of magnetic field at various temperatures shown in the legend, determined from data in Fig. 2 using Eqn. 2 as described in text. (c) Experimental $\rho_{xx}(H)$ (solid lines) and predicted $\rho_{xx}(H)$ (dashed lines) using parameters from (a) and (b) at three representative temperatures as indicated in legend.

**Figure 5 | Gate and field dependence of Hall resistivity and magnetoresistivity of WTe$_2$.** The magnetic field dependence of $\rho_{xy}$ (a), $\rho_{xx}$ (b) and magnetoresistance ratio MR (c) at 2 K at various electrolytic gate voltages shown in the legend. The inset in (b)

shows $\rho_{xx}$ at low field. The blue dashed lines in (a), (b) and (c) are fits to the two-fluid model Eqns. (1) and (2) for $V_g = 3$ V as described in the text.

**Figure 6 | Temperature dependence of $R_H$ and $\rho_{xx}$ of WTe$_2$.** Temperature dependence of Hall coefficient ($R_H$) and $\rho_{xx}$ at $H = 14$ T.

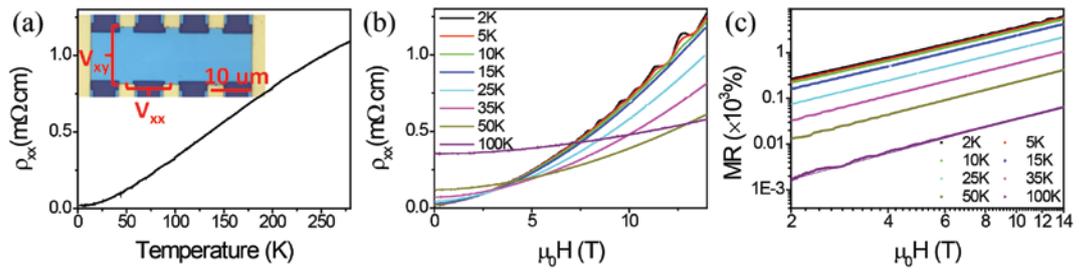

Y.L. Wang *et al*., Fig. 1

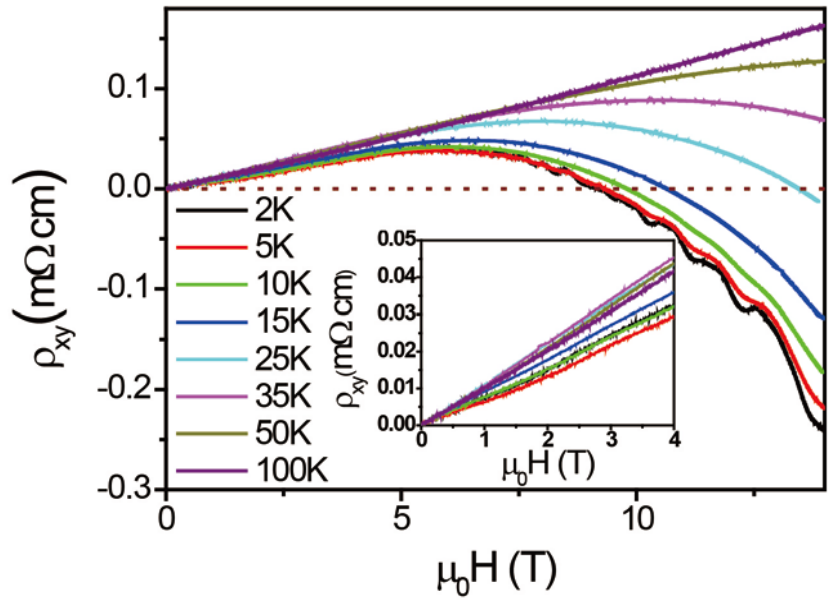

Y.L. Wang *et al.*, Fig. 2

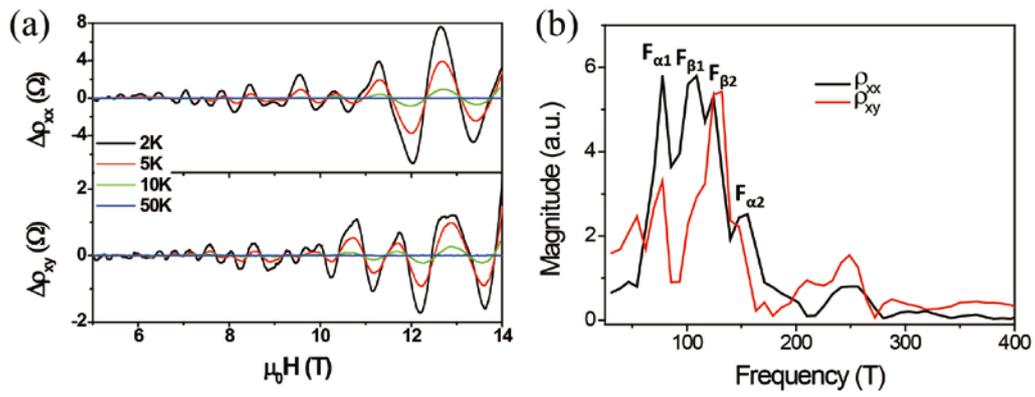

Y.L. Wang *et al.*, Fig. 3

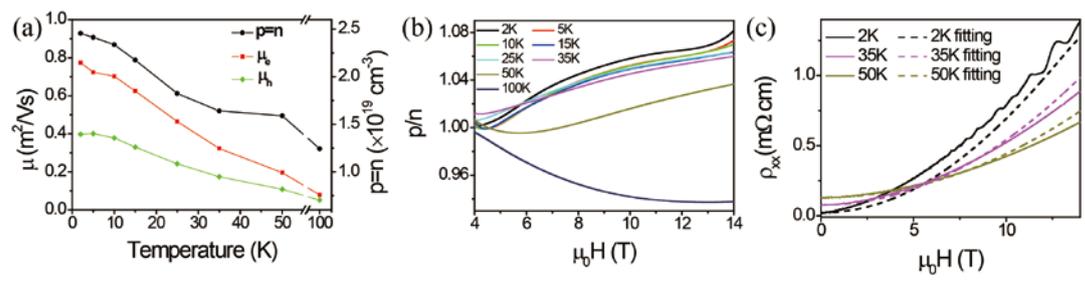

Y.L. Wang *et al.*, Fig. 4

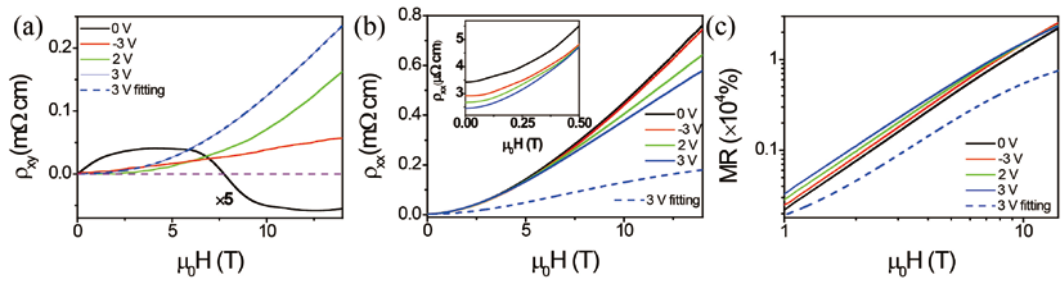

Y.L. Wang *et al.*, Fig. 5

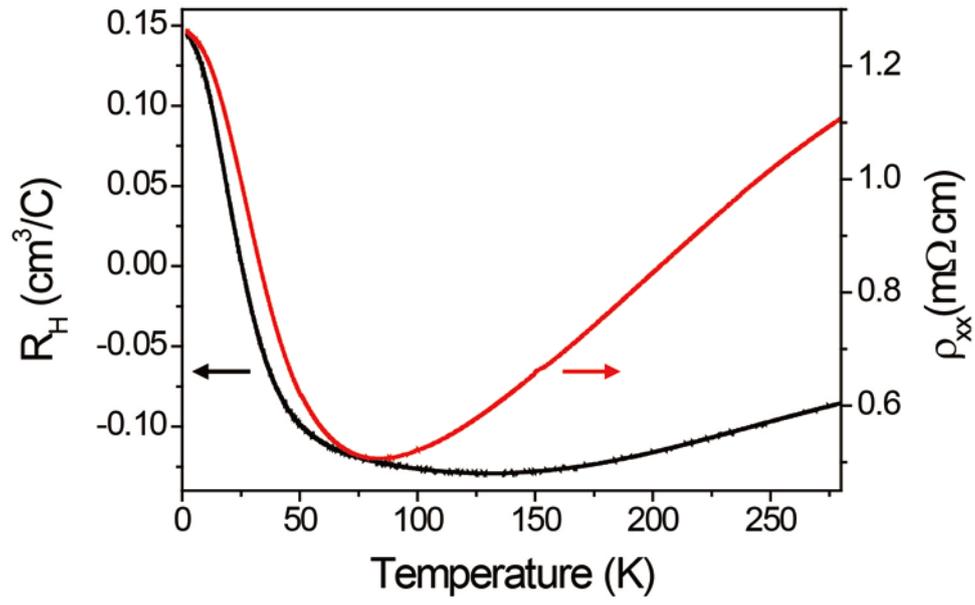

Y.L. Wang *et al*., Fig. 6